\documentclass[prl,twocolumn,showpacs,superscriptaddress]{revtex4}
\usepackage{epsfig,color}

\begin{document}
\newcommand{\ltwid}{\mathrel{\raise.3ex\hbox{$<$\kern-.75em\lower1ex\hbox{$\sim$}}}}
\newcommand{\gtwid}{\mathrel{\raise.3ex\hbox{$>$\kern-.75em\lower1ex\hbox{$\sim$}}}}
\newcommand{\bra}{\langle}
\newcommand{\ket}{\rangle}
\newcommand{\sill}{\psi}
\newcommand{\trace}{{\rm Tr}}
\newcommand{\ntilde}{\tilde{n}}
\newcommand{\stilde}{\tilde{s}}
\newcommand{\atilde}{\tilde{\alpha}}
\newcommand{\greg}[1]{\textcolor{red}{#1}}
\newcommand{\adrian}[1]{\textcolor{blue}{#1}}
\newcommand{\question}[1]{\textcolor{green}{#1}}

\title{Spin-incoherent behavior in the ground state of strongly correlated systems}

\author{Adrian E. Feiguin}
\affiliation{Department of Physics and Astronomy, University of Wyoming, Laramie, Wyoming 82071, USA}

\author{Gregory A. Fiete}
\affiliation{Department of Physics, The University of Texas at Austin, Austin, Texas 78712, USA}

\date{\today}

\begin{abstract}
It is commonly believed that strongly interacting one-dimensional Fermi systems with gapless excitations are effectively described by Luttinger liquid theory.  However, when the temperature of the system is high compared to the spin energy, but small compared to the charge energy, the system becomes ``spin-incoherent''. We present numerical evidence showing that the one-dimensional ``$t$-$J$-Kondo'' lattice, consisting of a $t$-$J$ chain interacting with localized spins, displays all the characteristic signatures of spin-incoherent physics, but in the {\it ground state}. 
We argue that similar physics may be present in a wide range of strongly interacting systems. \end{abstract}

\maketitle

The physics of interacting one-dimensional (1-d) fermionic systems is described by a universal effective theory called ``Luttinger liquid"  (LL) theory \cite{GiamarchiBook}, in which
  the low-energy physics is dominated by {\it bosonic} collective excitations.  The original fermions lose their identities as low-energy excitations, giving rise to the phenomenon of spin-charge separation, with distinct collective spin and charge excitations (spinons and holons, respectively) that have their own characteristic velocity and Hamiltonian. 


Recently, a previously overlooked regime at finite temperature has come to
light--the ``spin-incoherent Luttinger liquid" (SILL) \cite{Fiete2007b,Halperin2007}.
If the temperature is higher than the characteristic spin energy scale, but much smaller than the Fermi energy \cite{SILL_exp}, spins become totally incoherent, effectively at infinite temperature,
while the charge sector remains very close to its ground state.
 This regime is characterized by universal properties in the transport, tunneling density of states, and the spectral functions \cite{Fiete2007b}.
 

In an earlier work \cite{Feiguin2009d}, we have described the spectral
properties of a 1-d $t$-$J$ chain at finite temperature (corresponding to the strong coupling limit of the Hubbard model), and understood the
crossover from spin-coherent to spin-incoherent regimes in terms of a
transfer of spectral weight.  In this work we establish an analogy between finite temperature SILL physics, and the ground state properties of certain model Hamiltonians. 

\begin{figure}[b]
\centering
\epsfig{file=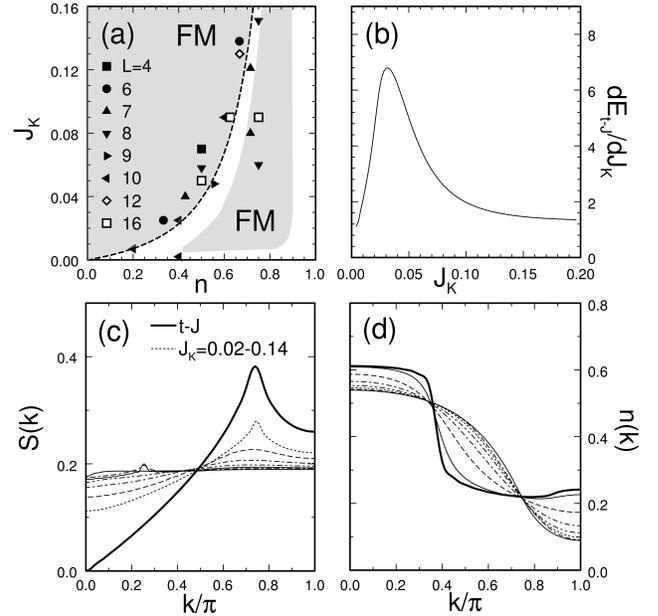,height=0.47\textwidth,angle=0}
\caption{
(a) Proposed phase diagram of the $t$-$J$-Kondo model with $J=0.05$, as a function of the density and Kondo coupling $J_K$, obtained with exact diagonalization on small systems (full symbols) and DMRG (open symbols). The gray shading corresponds to the FM phase, while the empty region is PM. The dashed line is a guide to the eye. (b) Derivative of $\langle H_{t-J}\rangle$ with respect to the Kondo coupling $J_K$, as explained in the text, for a chain with $L=32$, $N=24$ fermions, and $S^z=4$. (c) spin structure factor, and (d) momentum distribution of a $t$-$J$-Kondo chain with $L=64$ sites and $N=48$ fermions, for different values of the Kondo coupling $J_K$. 
}
\label{fig:gs}
\end{figure}

We motivate our results by establishing an analogy between (i) a thermal mixed state and (ii) a pure state in an enlarged Hilbert space.  This is the key idea behind the so-called thermo field formalism \cite{Takahashi1975}.
For illustration purposes, let us first assume that we have two spins $S=1/2$, that we put  into
a maximally entangled state
\begin{equation}
|I_0\rangle = \frac{1}{\sqrt{2}}
\left[|\uparrow,\tilde\downarrow\rangle \pm |\downarrow,\tilde\uparrow\rangle\right],
\label{epr}
\end{equation}
where the sign is irrelevant in the following treatment. We shall assume the first spin is our ``physical'' spin, while the one with a tilde is the ``ancilla'', or impurity spin. It is straightforward to see that the reduced density matrix of the physical spin, after tracing over the ancillary degrees of freedom, is the identity matrix.  Thus, if we assume that the ancilla acts as some sort of effective thermal bath, the physical spin is at infinite temperature.
It is easy to see that the maximally mixed state for a number of spins can be rewritten as:
$|I \rangle = \prod_i |I_{0i}\rangle,$
defining the maximally entangled state $|I_{0i}\rangle$ of spin $i$ with
its ``ancilla'', as in Eq.(\ref{epr}).
This construction allows one to represent a mixed state of a quantum system as a pure state in an enlarged Hilbert space 
and lies at the core of the imaginary-time DMRG \cite{Feiguin2005a}.

With this picture in mind, it is natural to draw an analogy to the physics of one-dimensional Kondo lattices \cite{Tsunetsugu1997}.  
Let us first consider a particular model describing a one dimensional chain of fermions with strong repulsive on-site interaction $U$, the $t$-$J$ model:
\begin{equation}
H_{t-J}=-t \sum_{i=1,\sigma}^L \left(c^\dagger_{i\sigma} c_{i+1\sigma}+\mathrm{h.c.}\right)
+ J \sum_{i=1}^L (\vec{s}_i \cdot \vec{s}_{i+1} -\frac{1}{4} n_i n_{i+1} ),
\label{H_t-J}
\end{equation}
with the implicit constraint forbidding double-occupancy. Here, $c^\dagger_{i\sigma}$ creates an electron of spin $\sigma$ on the
$i^{\rm th}$ site along a chain of length $L$. The exchange energy $J \sim t^2/U$,  and we take the inter-atomic distance as unity. We express all energies in units of the hopping parameter $t$.

In the $J=0$ limit, the ground state of the $t$-$J$ model
factorizes into the Ogata-Shiba wave-function \cite{ogata-shiba},
a product of a fermionic wave function $|\phi\rangle$, and a spin wave function
$|\chi\rangle$
\begin{equation}
|\mathrm{g.s.}\rangle=|\phi\rangle\otimes |\chi\rangle.
\label{gs}
\end{equation}
The first piece, $|\phi\rangle$, describes the charge degrees of freedom, and is
simply the ground state of a spinless non-interacting tight-binding Hamiltonian.
In this limit, the spin states are degenerate and the dispersion
is just a non-interacting band $\epsilon(k)=-2t\cos(k)$, but any finite interaction will lift this degeneracy and give the spin degree of freedom some dispersion.
Fig.1 in Ref.[\onlinecite{Feiguin2009d}] shows the spectrum for $J_K=0$ for a particular choice of parameters, $J/t=0.05$. 

Now it is easy to construct a generalization of the Ogata-Shiba wave-function to describe the system at {\it infinite spin temperature}. All we have to do is to add spin ancillas, and replace the spin component in Eq.(\ref{gs}), by the corresponding maximally entangled state, $|I \rangle = \prod_i |I_{0i}\rangle$, $|\psi_{SILL}\rangle=|\phi\rangle \otimes |I\rangle$. Thus, the charge will remain at zero temperature, while the spin component will be effectively at infinite temperature! This state is describing the spin-incoherent regime.

Now, leaving the ancillas aside for a moment, let us return to the original model and construct the full ``$t$-$J$-Kondo'' Hamiltonian by adding localized impurities interacting with the conduction fermions via an antiferromagnetic exchange, $J_K$:
\begin{equation}
H=H_{t-J}+J_K \sum_{i=1}^L \vec{s}_i \cdot \vec{S}_{i},
\label{fullH}
\end{equation}
where  $\vec{s}_i$ describes the conduction spins  and $\vec{S}_i$ the localized spins with $H_{t-J}$ the 1-d version of (\ref{H_t-J}).

Curiously, the ``$t$-$J$-Kondo'' lattice has not received much attention in the literature \cite{Moukouri1996}.
The ``Kondo-Hubbard'' lattice (from which the ``$t$-$J$-Kondo''  model can be derived)  has been studied in Ref. \cite{Yanagisawa1994}.
A key result is that its ground state for large Coulomb repulsion $U$ has a total spin $S=(L-N)/2$, where $L$ is the length (number of sites) of the chain, and $N$ is the total number of conduction fermions. This is similar to the situation for large $J_K$, and no Coulomb repulsion \cite{FM-Kondo, Tsunetsugu1997}. However, Coulomb repulsion can drive the system into a ferromagnetic (FM) ground state, even for small $J_K$ \cite{Moukouri1996}.

In the limit of large $J_K$ the fermions are strongly entangled with the localized spins, and the excitations become heavy polarons \cite{Sigrist1991,Smerat2009}. We are interested in the small $J$ (large $U$) regime where the spinons are almost dispersionless, and the coupling $J_K$ is small and of the order of $J$. In this case, the interaction $J_K$ is nominally a small perturbation, and one expects that the charge of the conduction fermions will not be affected since it is practically decoupled from the spin.
Let us assume first that $J=0$: an infinitesimal $J_K$ will pair the conduction spins to the impurities. For sufficiently large $J_K \gg t$ the corresponding state can be described by a product wave-function \cite{FM-Kondo}:
\begin{equation}
|\mathrm{g.s.}\rangle=|\phi\rangle\otimes |I\rangle \otimes |\sigma\rangle,
\label{gs Kondo}
\end{equation}
where the charge component $|\phi\rangle$ corresponds to pairs moving in a background of unpaired polarized impurities $|\sigma\rangle$. The pairs have their conduction spin maximally entangled to their impurity partner $|I\rangle$. By looking at the left two terms of this wave-function, we can easily identify the spin-incoherent state $|\psi_{SILL}\rangle$. Indeed, the unpaired impurities do not play a role in the dynamics of the conduction fermions. Thus, we have established a rigorous analogy between the ground state of the $t$-$J$-Kondo lattice in the $J=0$ limit, and the spin-incoherent state described by the Ogata-Shiba wave function at infinite spin temperature. Notice, however, that the charge excitation will have a gap for breaking a pair, that can be exponentially small for small $J_K$.

In the $J=0$ limit, the system is FM for any finite value of $J_K$. However, for a finite value of $J$, one expects that as the Kondo interaction $J_K$ is turned on, a paramagnetic (PM) window will open. In Fig.\ref{fig:gs}(a) we show a schematic phase diagram of the model for $J=0.05$, as a function of the density $n$ and Kondo coupling $J_K$. The data points correspond to the transition from a PM state with $S_{tot}=0$ to a state with finite $S_{tot}$, calculated with DMRG and exact diagonalization on small systems with open boundary conditions. Interestingly, our results suggest that the PM phase occupies a small sliver separating two large FM regions. The study of this phase diagram deserves further attention but remains out of the scope of this work.

We expect that the ground-state in the FM region at large $J_K$ will approximately be described by Eq.(\ref{gs Kondo}), while in the small $J_K$ region the system will be in a crossover regime:
the localized spins will act as an effective spin and thermal reservoir, by increasing the ``spin temperature'' of the conduction fermions and driving them spin-incoherent,  {\it in the ground state}.  Thus, the interaction $J_K$ parametrizes an effective temperature for the conduction electron spins.  We note that this ``temperature'' will not necessarily have a one-to-one correspondence with an actual temperature since our Hamiltonian is local.

To illustrate this point we have calculated the ground state for the Hamiltonian (\ref{fullH}) using the DMRG method \cite{White1992}.   We have picked the parameters $J=0.05$, and $n=N/L=0.75$ to be able to compare with the finite-temperature results of Ref.\cite{Feiguin2009d}.
As a technical point, we remark that unless othewise stated we work in the $S^z=0$ subspace.  Even though we have encountered a paramagnetic window for $J_K \approx 2J$ (see Fig. 1) \cite{Moukouri1996}, this is not relevant to our interpretation of the results, as we shall see below.

For $J_K=0$ the ground state is massively degenerate, since all configurations of localized spins will have the same energy. As we increase $J_K$, this degeneracy is lifted, but convergence is extremely difficult.  In our calculations we have retained up to $600$ states. 
Once the ground state wave function is determined, we can calculate the energy of the conduction fermions as $E_{t-J}  = \langle H_{t-J} \rangle$, where $H_{t-J}$ is described by Eq.(\ref{H_t-J}). We then define an effective ``specific heat'' as the derivative of the energy with respect to our effective temperature $J_K$, $C_{J_K}\equiv dE_{t-J}/dJ_K$, as shown in Fig.\ref{fig:gs}(b). In this case we have plotted the results for $L=32$, $N=24$, and $S^z=(L-N)/2=4$ to avoid level crossings. We point out that the results for $S^z=0$ are practically indistinguishable, meaning that the total spin does not seem to affect the general behavior of the conduction fermions.  These results should be compared to those obtained in \cite{Feiguin2009d} using time-dependent DMRG in imaginary time \cite{Feiguin2005a}. The features are qualitatively the same: The flattened curve after the hump indicates that all the spins have been thermally excited, and have basically thermalized at a value of $J_K \approx J$. It is noteworthy that this behavior reflecting a crossover between two different regimes remains hidden in the total energy and only becomes apparent after the $t$-$J$ contribution is taken into account separately.

\begin{figure} 
\centering
\epsfig{file=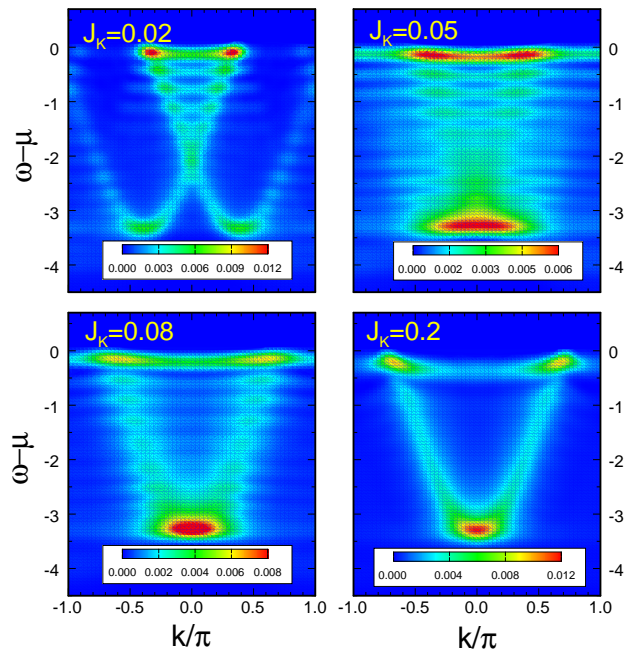,width=0.47\textwidth,angle=0}
\caption{
Momentum resolved spectrum of a $t-J$-Kondo chain of $L=32$ sites, with $J/t=0.05$ and density $n=0.75$, calculated with time-dependent density matrix renormalization group method, showing a transfer of the spectral weight as $J_K$ increases, toward a spectrum that resembles a spin-incoherent Luttinger liquid at a crossover value of $J_K \sim 0.05$.}
\label{fig:ak_kondo}
\end{figure}

To explore the analogy to spin-incoherent systems further, we have calculated the static spin structure factor $S(k)$, and the momentum distribution function $n(k)$ for the conduction fermions, as a function of $J_K$. In Fig.\ref{fig:gs}(c,d) we show our zero-temperature results for $L=64$,$N=48$, that should also be compared with those obtained at finite temperatures in \cite{Feiguin2009d}. Shown is the diagonal component of the spin-structure factor.  Again, the behavior is very similar and the analogy is clear: In $S(k)$ we see a peak at $k=2k_F$, with $k_F\equiv \pi N/2L$, and a pronounced minimum at $k=0$. For small values of $J_K$ the magnitude of the peak decreases and the minimum at $k=0$ increases. At temperatures of the order of $T \simeq J_K \simeq J$, the spin structure factor becomes essentially featureless, indicating that the spins are effectively at ``infinite'' temperature. These features are also observed in the conventional Kondo lattice, but in that case it corresponds to the formation of singlets in the strong $J_K$ regime \cite{Moukouri1995}. Note that for the range of parameters corresponding to the PM regime in Fig.1, a very small feature appears at $k=\pi(1-n)$, possibly indicating a large Fermi surface.

The behavior of $n(k)$ is even more enlightening. Again we find the same features observed in a finite temperature $t$-$J$ model:  At small values of $J_K$ we see the typical LL profile, with no discontinuities at the Fermi point and a singularity at $k=3k_F$. We also notice that the values $n(k_F)$ and $n(2k_F)$ are independent of $J_K$ within the accuracy of our calculation, as observed in Ref.\cite{Feiguin2009d} (where they are also independent of temperature). We see that the inflection point in $n(k)$ shifts from $k_F$ towards $2k_F$, indicating the onset of the spin-incoherent regime, understood as a shift from particles with spin dynamics to particles that are effectively spinless \cite{Cheianov2005}. 
This behavior resembles the physics of the Kondo lattice model, where the Fermi surface is enlarged by absorbing the local moments $\vec S_i$ into the Fermi sea. However, it differs from it in that this leads to a shift from $k_F= \pi N/2L$ to a new value of $\pi (N+L)/2L$ \cite{Oshikawa:prl97}, different than the value of $2k_F$ that occurs generically in the spin-incoherent regime.  
We believe that the large Fermi surface singularity is not seen due to the fact that system is dominated by FM correlations and spin-incoherent physics.

To confirm that the features observed above indeed correspond to a spin-incoherent regime at zero temperature, we have calculated the photoemission spectrum of the model using the time-dependent DMRG method as described in Ref.\cite{White2004a}. We have used the same parameters as in Ref.\cite{Feiguin2009d} for the finite temperature calculations. Our results for different values of $J_K$ are shown in Fig.\ref{fig:ak_kondo}. These spectra should be compared to those of the $t$-$J$ chain at finite temperature \cite{Feiguin2009d}. We can see a remarkable correspondence between the finite-temperature spectra, and the spectra of the $t$-$J$-Kondo chain. We first notice a lack of spectral weight above the Fermi level (whereas it is there in Ref. \cite{Feiguin2009d}), which is to be expected since these are zero-temperature calculations. We do not see an important change in the bandwidth, which means that for these values of $J_K$, the interaction with the localized spins has a minimum effect on the fermion effective mass. The most noteworthy feature is a transfer of spectral weight from the holon and shadow bands in such a way that at higher values of $J_K$ the spectrum resembles the dispersion for spinless fermions. This is precisely the behavior expected from a spin-incoherent Luttinger liquid, and is reinforced in the FM regime at $J_K=0.2$. The apparent discretization of the spectrum appears as a combination of two effects: the convolution of the holon dispersion with the relatively flat spinon dispersion, and the relatively small size of the system considered here.  One may argue that since there is a level crossing at finite $J_K$, these results may correspond to ground states with different excitations.
However, we have repeated the calculations for different $S^z$ subspaces, always finding similar behavior, irrespective of the ground state spin sector. Therefore, we can assert with confidence that these observations apply generically to the model.

To summarize, we have presented numerical results supporting our argument that in the small $J$ regime, the $t$-$J$-Kondo lattice may indeed display spin-incoherent behavior in the ground state, with the interaction $J_K$ parametrizing an effective temperature and the localized spins acting as an effective thermal bath.
 
We believe that this behavior may be a generic feature of many quasi-1D strongly interacting system systems, such as $t$-$J$ ladders. 
Preliminary results support this assertion and will be presented elsewhere \cite{t-J}.
 We can extrapolate our argument to higher dimensions, as well as heavy-fermion and multi-band systems \cite{Al-Hassanieh2009}: The observation of spin-charge separation in bulk systems, if present, may be hindered by small interactions that may wash out the characteristic signatures of the spin degrees of freedom.  Instead, one might only be able to see the charge excitations, with spectral properties that would resemble a gas of spinless fermions with a large Fermi surface. Therefore, experimental efforts seeking evidence of spin-charge separation may be more effectively focused toward looking for evidence of {\it spin-incoherent} behavior.

We gratefully acknowledge financial support from ARO grant W911NF-09-1-0527, NSF grant DMR-0955778 (G.A.F.), and from NSF grant DMR-0955707 (A.E.F.).



\end{document}